\makeatletter
\def\input@path{{sty/}}
\makeatother

\documentclass{egpubl}
\usepackage{eg2026}
 
\Poster                 % uncomment for Poster contribution
\usepackage[T1]{fontenc}
\usepackage{dfadobe}  
\usepackage{cite}  % comment out for biblatex with backend=biber
\usepackage{censor}
\BibtexOrBiblatex
\electronicVersion
\PrintedOrElectronic
\ifpdf \usepackage[pdftex]{graphicx} \pdfcompresslevel=9
\else \usepackage[dvips]{graphicx} \fi
\usepackage{svg}        % for SVG images

\usepackage{egweblnk} 
\usepackage{amsmath}    % align, cases, etc.
\usepackage{amssymb}    % symbols
\usepackage{bm}         % bold math symbols
\usepackage{dsfont}     % indicator ��

\usepackage[dvipsnames]{xcolor}

\usepackage{booktabs}   % professional tables
\usepackage{multirow}
\usepackage{makecell}
\usepackage{colortbl}   % row/column coloring

\usepackage{algorithm}
\usepackage{algpseudocode}

\usepackage{siunitx}

\usepackage{tikz}
\usetikzlibrary{
  arrows.meta,
  positioning,
  calc,
  shapes.geometric,
  decorations.pathreplacing,
  decorations.pathmorphing,
  backgrounds,
  fit,
  patterns,
  fadings
}

\usepackage{pgfplots}
\pgfplotsset{compat=1.18}
\usepgfplotslibrary{groupplots,colormaps}

\usepackage{tcolorbox}

\usepackage{pifont}

\definecolor{agentblue}{RGB}{70,130,180}
\definecolor{targetorange}{RGB}{230,120,50}
\definecolor{selectgreen}{RGB}{60,160,80}
\definecolor{warnred}{RGB}{200,80,80}

\definecolor{panelbg}{RGB}{252,252,252}
\definecolor{headerbg}{RGB}{235,235,240}

\definecolor{viridispurple}{HTML}{440154}
\definecolor{viridisyellow}{HTML}{E7E41E}

\definecolor{oursPink}{HTML}{FFB5A7}
\definecolor{oursGreen}{HTML}{CAFFBF}
\definecolor{spotBlue}{RGB}{220,235,255}

\definecolor{Seaborn1}{RGB}{103,136,238}
\definecolor{Seaborn2}{RGB}{154,187,255}
\definecolor{Seaborn3}{RGB}{201,215,240}
\definecolor{Seaborn4}{RGB}{237,209,194}
\definecolor{Seaborn5}{RGB}{247,168,137}
\definecolor{Seaborn6}{RGB}{226,105,82}

\tikzfading[
  name=fade along path,
  left color=transparent!70,
  right color=transparent!0
]

\pgfdeclareplotmark{hexagon}{
  \pgfpathmoveto{\pgfqpoint{0pt}{2pt}}
  \pgfpathlineto{\pgfqpoint{1.73pt}{1pt}}
  \pgfpathlineto{\pgfqpoint{1.73pt}{-1pt}}
  \pgfpathlineto{\pgfqpoint{0pt}{-2pt}}
  \pgfpathlineto{\pgfqpoint{-1.73pt}{-1pt}}
  \pgfpathlineto{\pgfqpoint{-1.73pt}{1pt}}
  \pgfpathclose
  \pgfusepathqfillstroke
}

\title[VoroFields]%
      {Neural Approximation of Generalized Voronoi Diagrams}

\author[P. Rigas, G. Ioannakis \& I. Emiris]
{\parbox{\textwidth}{\centering
Panagiotis Rigas$^{1,2}$\orcid{0009-0006-5395-6112},
George Ioannakis$^{3}$\orcid{0000-0001-6230-7030},
and Ioannis Emiris$^{1,2}$\orcid{0000-0002-2339-5303}
}
\\
{\parbox{\textwidth}{\centering
$^{1}$Department\ of Informatics \& Telecommunications, National and Kapodistrian University of Athens, Greece\\
$^{2}$Archimedes, Athena Research Center, Greece\\
$^{3}$Institute for Language and Speech Processing, Athena Research Center, Greece
}}
}

\begin{document}

% uncomment for using teaser
\teaser{
 \includegraphics[width=0.7\linewidth]{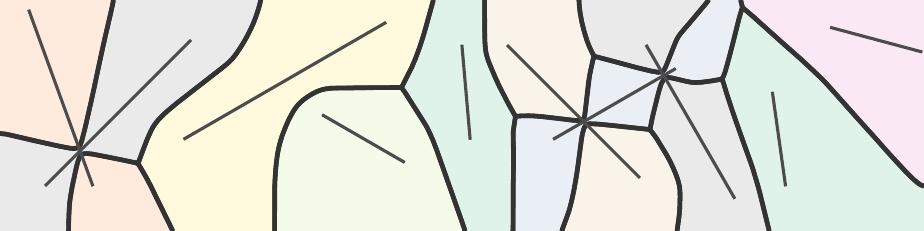}
 \centering
  \caption{Neural implicit Voronoi partition of line segments induced by learned score functions.}
\label{fig:teaser}
}

\maketitle
%-------------------------------------------------------------------------
\begin{abstract}

We introduce \emph{VoroFields}, a hierarchical neural-field framework for approximating generalized Voronoi diagrams of finite geometric site sets in low-dimensional domains under arbitrary evaluable point-to-site distances. Instead of constructing the diagram combinatorially, VoroFields learns a continuous, differentiable surrogate whose maximizer structure induces the partition implicitly. The Voronoi cells correspond to maximizer regions of the field, with boundaries defined by equal responses between competing sites. A hierarchical decomposition reduces the combinatorial complexity by refining only near envelope transition strata. Experiments across site families and metrics demonstrate accurate recovery of cells and boundary geometry without shape-specific constructions.
% We introduce \emph{VoroFields}, a hierarchical neural-field framework for approximating generalized Voronoi diagrams of finite sets of geometric sites in low-dimensional domains under arbitrary point-to-site distance functions. Rather than constructing the diagram explicitly, VoroFields learns a continuous surrogate whose maximizer structure induces the Voronoi partition implicitly. The diagram emerges as the dominance stratification of the neural field: cells correspond to regions of maximal response, and boundaries lie along loci of equal influence between competing sites. A hierarchical refinement scheme enables scalability to large site sets by concentrating resolution near geometrically complex regions. Experiments across site families and metrics demonstrate accurate recovery of cells and boundary structure. 

\vspace{4pt}
\end{abstract}  
%-------------------------------------------------------------------------
\section{Introduction}
\label{sec:intro}

\begin{figure}
    \centering
    \begin{tikzpicture}[scale=0.75]
% Input: 1D line segments, diff lengths
        \begin{scope}[xshift=2cm]
            \node[font=\small\bfseries] at (0,3) {1D Line Segments};
            
            % 1D line with segments of different lengths
            \draw[thick] (-1.2,1.5) -- (1.2,1.5);
            
            % Line segments with different lengths
            \draw[red!70,very thick,line cap=round] (-0.9,1.5) -- (-0.5,1.5);      % length 0.4
            \draw[teal!70,very thick,line cap=round] (-0.1,1.5) -- (0.3,1.5);       % length 0.4  
            \draw[orange!70,very thick,line cap=round] (0.6,1.5) -- (1.1,1.5);      % length 0.5
            
            % Endpoints
            \fill[red!70] (-0.9,1.5) circle (0.04);
            \fill[red!70] (-0.5,1.5) circle (0.04);
            \fill[teal!70] (-0.1,1.5) circle (0.04);
            \fill[teal!70] (0.3,1.5) circle (0.04);
            \fill[orange!70] (0.6,1.5) circle (0.04);
            \fill[orange!70] (1.1,1.5) circle (0.04);
            
            % Labels
            \node[red!70,below] at (-0.7,1.4) {\(s_1\)};
            \node[teal!70,below] at (0.1,1.4) {\(s_2\)};
            \node[orange!70,below] at (0.85,1.4) {\(s_3\)};
            
            \node[font=\footnotesize] at (0,0.8-0.1) {Segments in \(\mathbb{R}^1\)};
        \end{scope}
        
        % Arrow
        \draw[thick,->] (4.5,1.5) -- (5.5,1.5);
        \node[above] at (5,1.8) {\footnotesize \(f_e(x) = d(x, s_e)\)};
        
        % Function arrangement (2D graph) - no y-axis
        \begin{scope}[xshift=8cm]
            \node[font=\small\bfseries] at (0,3) {Distance Functions};
            
            % Piecewise linear distance functions for segments
            % Segment 1: flat on [-0.9,-0.5], linear outside
            \draw[red!70,thick] (-1.2,1.7) -- (-0.9,1.4) -- (-0.5,1.4) -- (0.4,2.3);
            
            % Segment 2: flat on [-0.1,0.3], linear outside  
            \draw[teal!70,thick] (-1.,2.3) -- (-0.1,1.4) -- (0.3,1.4) -- (1.,2.3);
            
            % Segment 3: flat on [0.6,1.1], linear outside
            \draw[orange!70,thick] (-0.3,2.3) -- (0.6,1.4) -- (1.1,1.4) -- (1.2,1.5);
            
            % Function labels
            \node[red!70] at (-0.9,1.8+0.025) {\(f_1\)};
            \node[teal!70] at (-0.45,2.15+0.025) {\(f_2\)};
            \node[orange!70] at (1.0,1.7-0.05) {\(f_3\)};
            
            % Only x-axis
            \draw[->] (-1.3,1.2) -- (1.3,1.2) node[right] {\tiny \(x\)};
            
            \node[font=\footnotesize] at (0,0.8-0.1) {Arrangement in \(\mathbb{R}^1 \times \mathbb{R}\)};
        \end{scope}
        
        % Arrow
        \draw[thick,->] (8,0.3) -- (8,-0.7);
        \node[right] at (8.5,-0.2) {\footnotesize \(\min\)};
        
        % Lower envelope - no y-axis
        \begin{scope}[xshift=8cm, yshift=-3cm]
            \node[font=\small\bfseries] at (0,0) {Lower Envelope};
            
            % Correct lower envelope
            \draw[very thick,red!80] (-1.2,1.7) -- (-0.9,1.4) -- (-0.5,1.4) -- (-0.3, 1.6);
            \draw[very thick,teal!80] (-0.3, 1.6) -- (-0.1, 1.4)--(0.3, 1.4)--(0.45, 1.55);
            \draw[very thick, orange!80] (0.45, 1.55) -- (0.6,1.4) -- (1.1,1.4) -- (1.2,1.5);
            % Key intersection points
            \fill[black] (-0.3, 1.6) circle (0.04);
            \fill[black] (0.45, 1.55) circle (0.04);
            
            % Only x-axis
            \draw[->] (-1.3,1.2) -- (1.3,1.2) node[right] {\tiny \(x\)};
            
            \node[font=\footnotesize] at (0,0.8-0.1) {\(L(x) = \min_{e\in E} f_e(x)\)};
        \end{scope}
        
        % Arrow down
        \draw[thick,->] (5.5,-1.5) -- (4.5,-1.5);
        \node[above] at (5,-1.2) {\footnotesize project};
        
        % Correct Voronoi diagram (1D)
        \begin{scope}[xshift=2cm,yshift=-3cm]
            \node[font=\small\bfseries] at (0,0) {1D Voronoi};
            
            % Correct 1D Voronoi regions based on lower envelope
            \draw[very thick,red!70] (-1.2,1.4) -- (-0.3,1.4);      % s1 wins from left to first intersection
            \draw[very thick,teal!70] (-0.3,1.4) -- (0.45,1.4);     % transition region
            \draw[very thick,orange!70] (0.45,1.4) -- (1.2,1.4);     % s3 wins on right
            
            % Key intersection points
            \fill[black] (-0.3, 1.4) circle (0.04);
            \fill[black] (0.45, 1.4) circle (0.04);
            
            % Original segments at bottom
            \draw[red!70,very thick,line cap=round] (-0.9, 1.2) -- (-0.5,1.2);
            \draw[teal!70,very thick,line cap=round] (-0.1,1.2) -- (0.3,1.2);
            \draw[orange!70,very thick,line cap=round] (0.6,1.2) -- (1.1,1.2);
            
            \node[font=\footnotesize] at (0, 0.8-0.1) {\(V_T = \{x \in D: \arg\min_{e\in E} f_e(x)=T\}\)};
        \end{scope}
\end{tikzpicture}
% }
    \caption{Arrangement of distance functions for 1D segments, its lower envelope, and the projected Voronoi stratification.}
    \label{fig:classic-framework}
    \vspace{-1.5mm}
\end{figure}
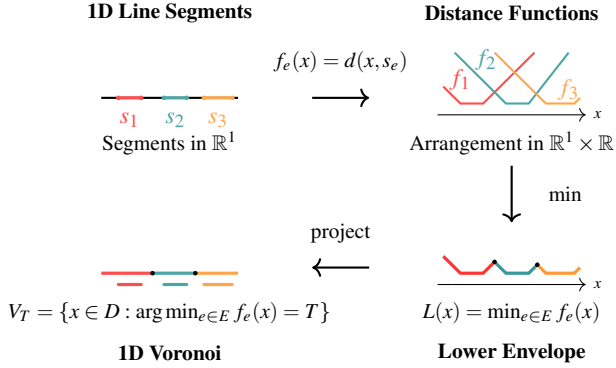

\textbf{Geometric Setting.} Let $D \subseteq \mathbb{R}^d$ be a domain and let 
$E=\{1,\dots,n\}$ index a finite family of sites 
$\{s_e\}_{e\in E}$. 
Each site induces a real-valued function 
$f_e : D \to \mathbb{R},\ x \mapsto f_e(x)$, typically a point-to-site distance.
The associated arrangement of graphs 
$\Gamma_e := \{(x,r)\in D\times\mathbb{R} : r=f_e(x)\}$ 
partitions $D\times\mathbb{R}$ into cells determined by the sign pattern of 
$\{r-f_e(x)\}_{e\in E}$~\cite{aurenhammer2013voronoi}. 
The generalized Voronoi stratification of $D$ is the projection of the $1$-level cells of this arrangement, equivalently the decomposition induced by the lower envelope
\begin{equation}
L(x):=\min_{e\in E} f_e(x),
\qquad
V_T:=\{x\in D:\arg\min_{e\in E} f_e(x)=T\},
\label{eq:loenv}
\end{equation}
for nonempty $T\subseteq E$. Figure~\ref{fig:classic-framework} illustrates the classical construction:
distance functions  \(f_e(x):=d(x,s_e) \in \mathbb{R}_{\geq 0}\), their graphs, the lower envelope, and the projected Voronoi stratification.

\noindent\textbf{Motivation.}
% Voronoi diagrams are lower envelopes of distance fields. 
% Classical constructions compute the envelope combinatorially \cite{aurenhammer2013voronoi}, with certified algebraic variants in 2D \cite{emiris2013voronoi} and $\varepsilon$-approximate or octree-based discretizations for points and 3D settings \cite{har2001replacement,boada2008approximations}. 
% We instead learn a continuous surrogate whose maximizer structure induces the stratification implicitly.
Exact constructions of minimization diagrams rely on geometric predicates tailored to specific site families and metrics, often requiring algebraic computation and certified robustness~\cite{aurenhammer2013voronoi,emiris2013voronoi}. 
Approximate approaches typically subdivide space (e.g., via quad/octrees), assign nearest-site labels to vertices or cells, and reconstruct the diagram from local topology tests~\cite{boada2008approximations}.
Such approaches approximate the distance field over a spatial data structure, and their memory and complexity scale with spatial resolution.
We instead approximate the lower-envelope operator itself by a continuous, differentiable surrogate, avoiding spatial discretization or explicit combinatorial construction, and enabling amortized evaluation when distance queries are repeated or expensive.

\noindent\textbf{Contributions.}
We approximate the envelope by a neural field~\cite{xie2022neural} $\Phi_\Theta : D \to \mathbb{R}^{n}$ and define the neural envelope
\begin{equation}
L_\Theta(x)=\max_{e\in E} \Phi_{\Theta,e}(x),
\quad
V_T=\{x\in D:\arg\max_{e\in E} \Phi_{\Theta,e}(x)=T\}.
\label{eq:neural_approx}
\end{equation}
This replaces spatial discretization by a continuous, differentiable, compressed, and implicit surrogate of the minimization operator. 
To scale with large $n$, we factor the global maximization into a hierarchy of local decisions over subsets $E_v\subset E$, $|E_v|=k\ll n$, yielding depth $h\approx\lceil\log_k n\rceil$, site-based (rather than space-based) refinement, and uniform handling of heterogeneous distance functions.

\begin{figure}
    \centering
    \includegraphics[width=0.96\linewidth]{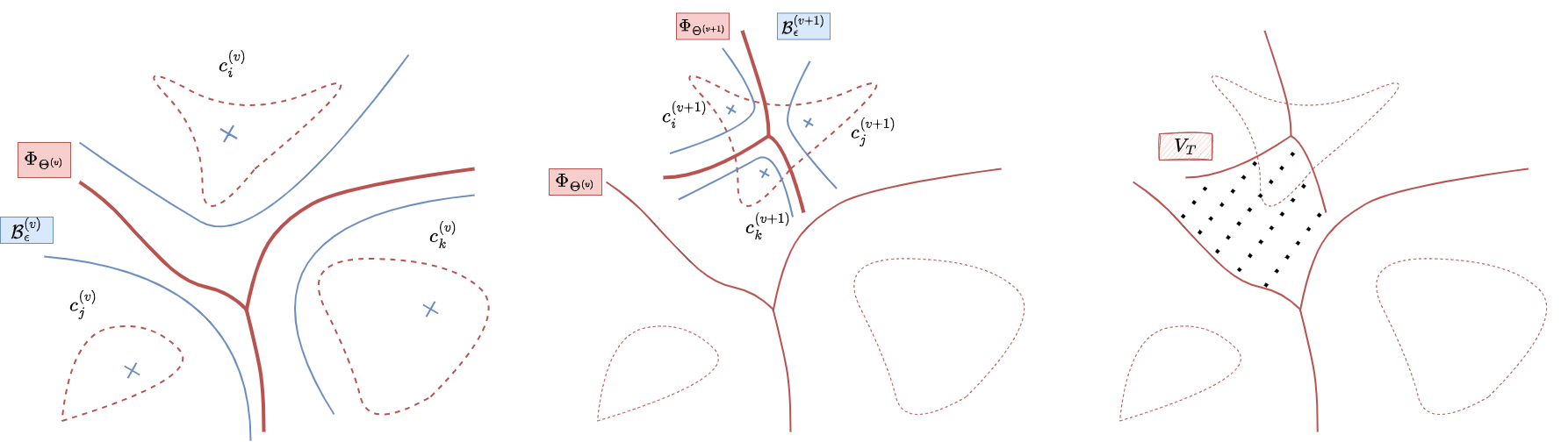}
    \caption{Hierarchical VoroFields construction. Local partitioning, fat-bisector sampling, and recursive refinement of Voronoi strata. Final regions are obtained via tree parsing.}
    \label{fig:method}
    \vspace{-1mm}
\end{figure}

\section{VoroFields}

We approximate the lower envelope $L$ in Eqt~\eqref{eq:loenv}
by learning a neural surrogate $\Phi_\Theta : D \to \mathbb{R}^{|E|}$.
The induced neural stratification is defined in Equation~\eqref{eq:neural_approx},
and training enforces
\(
\arg\max_{e\in E} \Phi_{\Theta,e}(x)
\approx
\arg\min_{e\in E} f_e(x).
\)

\noindent\textbf{Hierarchical Decomposition.}
To scale to large $n$, we construct a tree
\(
\mathcal{T}
=
\{(v,\Phi_{\Theta^{(v)}}, \mathcal{R}^{(v)}, E^{(v)})\}_{v\in\mathcal{V}},
\)
where each node $v$ contains a subset $E^{(v)}\subset E$,
a spatial region $\mathcal{R}^{(v)}\subset D$,
and a local neural field
$\Phi_{\Theta^{(v)}} : \mathcal{R}^{(v)} \to \mathbb{R}^{|E^{(v)}|}$.
Subsets are obtained via $k$-means clustering of sites,
yielding partitions
$E^{(v)} \to \{E^{(v)}_1,\dots,E^{(v)}_k\}$
with $|E^{(v)}_i|\ll |E^{(v)}|$.
Inference replaces the global maximization over $E$
by successive local ones along the tree
until a leaf $v_L$ is reached.

\noindent\textbf{Fat-Bisector Sampling.}
Within each node $v$, Voronoi interfaces correspond to
$\exists i \neq j \in E^{(v)} \ \text{s.t.}\
f_i(x)=f_j(x)=\min_{e\in E^{(v)}} f_e(x).$
% $f_i(x)=f_j(x)$ for $i,j\in E^{(v)}$.
We concentrate samples inside the $\varepsilon$-thickened boundary
\begin{equation}
\mathcal{B}_\varepsilon^{(v)}
:=
\left\{
x\in\mathcal{R}^{(v)} :
\exists i\neq j \text{ s.t. }
f_i(x),f_j(x)
\le
\min_{e} f_e(x)+\varepsilon
\right\}.
% \{ x\in \mathcal{R}^{(v)} :
% \min_{i\neq j \in E^{(v)}} |f_i(x)-f_j(x)| \le \varepsilon \}.
\label{eq:fat_bisector}
\end{equation}
In practice, for low dimensions, this is approximated using cluster centroids
$c_i^{(v)}$ and surrogate distances
$\tilde f_i^{(v)}(x)=\|x-c_i^{(v)}\|$.

\noindent\textbf{Supervision.}
For sampled $x\in D$, labels are computed as
$y(x)=\arg\min_{e\in E} f_e(x)$,
and parameters are optimized via
\(
\mathcal{L}(\Theta)
=
\mathbb{E}_{x\sim\mathcal{D}}
\big[
\ell(\Phi_\Theta(x), y(x))
\big],
\)
with $\ell$ the cross-entropy loss.
\section{Experimental Results}

We evaluate VoroFields on synthetic geometric site families with $n=10$k parametric sites per dataset (e.g.\ cuboids, ellipses, segments) in 2D and 3D, using analytic distances $f_e(x)$. Such non-point sites already require specialized nearest-site constructions even in low dimension \cite{abdelkader2021approximate}. 
High Top-1/Top-2 accuracy across dimensions, metrics, and primitives (Table~\ref{table:leaf_accuracy}) indicates reliable routing. 
Increasing segment length stresses boundary complexity; increasing $\varepsilon$ in Eqt~\eqref{eq:fat_bisector} improves resolution (Table~\ref{tab:boundary_ablation}). 
Errors concentrate near dense bisector intersections and depend on $\varepsilon$ and sampling density. 
Notably, training on the lower envelope (Top-1 $\approx0.9588$) yields representations partially consistent with higher-order structure ( Kendall--$\tau \approx 0.7215$, order-2 $\approx 0.7746$; evaluated on a 200-segment subset), suggesting that training through the probability simplex induces implicit ordering in the learned scores.
% We evaluate VoroFields on synthetic families with $n=10$k parametric sites per dataset (e.g.\ segments $(x,y,\ell,\theta)$, ellipses $(x,y,a,b,\theta)$), using analytic distances $f_e(x)$.  
% The hierarchy partitions $E$ via $k$-means until $|E^{(v)}|\le\tau$, giving depth $h \approx \lceil \log_k(n/\tau)\rceil$ (\emph{Layers}).  
% Accuracy is measured at the reached leaf by ranking candidates via $\Phi_{\Theta^{(v_L)}}(x)$ and reporting Top-$m$ success.

% We evaluate VoroFields on synthetic geometric site families with $n=10$k parametric sites per dataset (e.g.\ segments $(x,y,\ell,\theta)$, ellipses $(x,y,a,b,\theta)$), using analytic distances $f_e(x)$; such non-point sites already require specialized nearest-site constructions even in low dimension~\cite{abdelkader2021approximate}. Table~\ref{table:leaf_accuracy} shows high Top-1 / Top-2 accuracy across dimensions, metrics, and primitive families, indicating reliable routing.  
% Table~\ref{tab:boundary_ablation} increases segment length to stress boundary complexity: enlarging $\varepsilon$ in Equation~\eqref{eq:fat_bisector} improves boundary resolution. 
% Performance drops near densely intersecting bisectors, where accuracy depends on $\varepsilon$ and sampling density; extending to higher-order strata remains future work. Interestingly, training only on the lower envelope yields representations that are largely consistent with higher-order site orderings (Kendall--$\tau \approx 0.72$).

\begin{table}[t]
\centering
\resizebox{0.8\columnwidth}{!}{
\begin{tabular}{lccccc}
\toprule
Dataset & Layers & Dim & Metric & Top-1 (\%) & Top-2 (\%) \\
\midrule
Squares      & 5 & $\mathbb{R}^2$ & $L_\infty$ & $93.46 \pm 1.75$ & $97.53 \pm 0.29$ \\
Cuboids      & 7 & $\mathbb{R}^3$ & $L_\infty$ & $89.40 \pm 1.84$ & $97.46 \pm 0.83$ \\
Ellipses     & 5 & $\mathbb{R}^2$ & $L_2$      & $94.73 \pm 1.01$ & $98.46 \pm 0.56$ \\
Segments     & 5 & $\mathbb{R}^2$ & $L_2$      & $94.47 \pm 1.21$ & $97.93 \pm 0.49$ \\
Segments     & 7 & $\mathbb{R}^3$ & $L_2$      & $93.72 \pm 1.83$ & $97.59 \pm 1.09$ \\
NC-Polygons  & 7 & $\mathbb{R}^2$ & $L_2$      & $92.66 \pm 1.73$ & $97.20 \pm 0.37$ \\
\bottomrule
\end{tabular}
}
\caption{Leaf-level Top-$m$ accuracy (\%) across shape families ($n=10$k sites). Layers denote hierarchy depth until leaf size $\tau$.}
\label{table:leaf_accuracy}
\vspace{-.5mm}
\end{table}
\begin{table}[t]
\centering
\small
\resizebox{0.6\columnwidth}{!}{
\begin{tabular}{lccccc}
\toprule
$\ell_{\max}/\text{bbox}$ & 10\% & 20\% & 30\% & 40\% & 50\% \\
\midrule
Baseline                  & 91.93 & 84.20 & 68.46 & 71.13 & 72.80 \\
Increased $\varepsilon$   & 94.33 & 89.34 & 78.53 & 75.53 & 76.80 \\
\bottomrule
\end{tabular}
}
\caption{Top-1 accuracy (\%) for 2D segments under increasing relative length. Enlarging $\varepsilon$ improves boundary coverage.}
\label{tab:boundary_ablation}
\vspace{-1.5mm}
\end{table}
\section*{Acknowledgments}
This work has been partially supported by project MIS 5154714 of the National Recovery and Resilience Plan Greece 2.0 funded by the European Union under the NextGenerationEU Program.
%-------------------------------------------------------------------------
% \bibliography{egbibsample}
% \bibliographystyle{}
%-------------------------------------------------------------------------
% bibtex
\bibliographystyle{eg-alpha-doi} 
\bibliography{egbibsample}       

% biblatex with biber
% \printbibliography            
\end{document}